\documentclass[reprint,superscriptaddress, amsmath,amssymb,aps]{revtex4-1}

\usepackage{graphicx}
\usepackage{dcolumn}
\usepackage{bm}
\usepackage{hyperref}

\usepackage[dvipdfmx]{color}

\begin{document}

\title{Negative Energy Elasticity in a Rubberlike gel}
\author{Yuki~Yoshikawa}
\thanks{These authors contributed equally: Y. Yoshikawa, N. Sakumichi}
\affiliation{Department of Bioengineering, Graduate School of Engineering, The University of Tokyo, 7-3-1 Hongo, Bunkyo-ku, Tokyo, Japan.}
\author{Naoyuki~Sakumichi}
\thanks{These authors contributed equally: Y. Yoshikawa, N. Sakumichi}
\affiliation{Department of Bioengineering, Graduate School of Engineering, The University of Tokyo, 7-3-1 Hongo, Bunkyo-ku, Tokyo, Japan.}
\author{Ung-il~Chung}
\affiliation{Department of Bioengineering, Graduate School of Engineering, The University of Tokyo, 7-3-1 Hongo, Bunkyo-ku, Tokyo, Japan.}
\author{Takamasa~Sakai}
\email[Correspondence should be addressed to N.~Sakumichi or T.~Sakai:\\]{sakumichi@tetrapod.t.u-tokyo.ac.jp;\\
sakai@tetrapod.t.u-tokyo.ac.jp}
\affiliation{Department of Bioengineering, Graduate School of Engineering, The University of Tokyo, 7-3-1 Hongo, Bunkyo-ku, Tokyo, Japan.}
\date{\today}

\begin{abstract}
Rubber elasticity is the archetype of the entropic force emerging from the second law of thermodynamics; numerous experimental and theoretical studies on natural and synthetic rubbers have shown that the elasticity originates mostly from entropy change with deformation.
Similarly, in polymer gels containing a large amount of solvent, it has also been postulated that the shear modulus (the modulus of rigidity) $G$, which is a kind of modulus of elasticity, is approximately equivalent to the entropy contribution $G_S$, but this has yet to be verified experimentally.
In this study, we measure the temperature dependence of the shear modulus $G$ in a rubberlike (hyperelastic) polymer gel whose polymer volume fraction is at most 0.1. 
As a result, we find that the energy contribution $G_E=G-G_S$ can be a significant negative  value, reaching up to double the shear modulus $G$ (i.e., $\left|G_E\right| \simeq 2G$), although the shear modulus of stable materials is generally bound to be positive.
We further argue that the energy contribution $G_E$ is governed by a vanishing temperature that is a universal function of the normalized polymer concentration, and $G_E$ vanishes when the solvent is removed.
Our findings highlight the essential difference between rubber elasticity and gel elasticity (which were previously thought to be the same) and push the established field of gel elasticity into a new direction.
\end{abstract}

\maketitle

\section{Introduction}
\label{sec:Introduction}

Rubbers and rubberlike polymer gels are composed of three-dimensional networks of cross-linked polymer chains and have very different elastic properties compared to hard solids such as metals and ceramics: softness, high elongation, and an evident Gough-Joule effect.
This difference stems from the difference in the origin of elasticity.
The elastic properties of hard solids are explained by energy elasticity, which originates from internal energy changes resulting mainly from changes in bond angles and bond lengths.
On the other hand, the anomalous elastic properties of rubbers and rubberlike polymer gels are explained by entropy elasticity, which originates from entropy changes resulting mainly from changes in the conformation of polymer chains \cite{Flory1953, deGennes1979, Rubinstein2003}.\\

We can experimentally determine the entropy contribution $\sigma_S$ and the energy contributions $\sigma_E$ by measuring the (shear) stress $\sigma$ as a function of temperature $T$ under a constant-volume condition \cite{Flory1953, Rubinstein2003} 
(the van 't Hoff isochore \cite{Fermi1937}).
We consider an incompressible elastomer and apply a (shear) strain $\gamma$.
In an isothermal process, the corresponding stress 
$\sigma =\sigma (T,\gamma)$
is related to the Helmholtz free energy density $f=f(T,\gamma)$ as $\sigma = \partial f/\partial \gamma$.
On the basis of $f=e-Ts$, where $e$ is the internal energy density and $s$ is the entropy density, we can separate the entropy contribution $\sigma_S$ and the energy contribution $\sigma_E$ to the stress $\sigma=\sigma_S+\sigma_E$ as
 $\sigma_{S} \equiv  - T\partial s/\partial \gamma$ and $\sigma_{E} \equiv \partial e/\partial \gamma$.
According to the Maxwell relation $\partial s/\partial \gamma = -\partial \sigma/\partial T$, we have
\begin{equation}
\sigma_{S} (T,\gamma)=T\frac{\partial \sigma}{\partial T} (T,\gamma).
\label{eq:sigma-vantHoff}
\end{equation}
Using Eq.~(\ref{eq:sigma-vantHoff}), we can determine $\sigma_{S}$ by measuring the $T$ dependence of $\sigma$ when $\gamma$ is fixed.
Then, we can obtain $\sigma_{E}$ as $\sigma_{E}=\sigma-\sigma_{S}$.
We note that $\sigma_{S}$ and $\sigma_{E}$ are defined under a constant-volume condition.\\

Similarly, we can experimentally determine the entropy contribution $G_S$ and the energy contributions $G_E$ by measuring the shear modulus (the modulus of rigidity) $G$ as a function of temperature $T$ under a constant-volume condition.
Here, the shear modulus is defined by
\begin{equation}
G(T)\equiv \left. \frac{\partial \sigma}{\partial \gamma} (T,\gamma) \right|_{\gamma =0}
= \left. \frac{\partial^{2} f}{\partial \gamma^{2}} (T,\gamma) \right|_{\gamma =0}.
\label{eq:def:G}
\end{equation}
The entropy and energy contributions to the shear modulus are defined by
\begin{equation}
G_S(T)\equiv \left. \frac{\partial \sigma_S}{\partial \gamma} (T,\gamma) \right|_{\gamma =0}
= \left. -T\frac{\partial^{2} s}{\partial \gamma^{2}} (T,\gamma) \right|_{\gamma =0}
\label{eq:def:GS}
\end{equation}
and
\begin{equation}
G_E(T)\equiv \left. \frac{\partial \sigma_E}{\partial \gamma} (T,\gamma) \right|_{\gamma =0}
= \left. \frac{\partial^{2} e}{\partial \gamma^{2}} (T,\gamma) \right|_{\gamma =0},
\label{eq:def:GE}
\end{equation}
respectively.
In the same manner as in Eq.~(\ref{eq:sigma-vantHoff}), we can determine the entropy and energy contributions as
\begin{align}
&G_{S}(T)=T\frac{dG}{dT}(T),\\
&G_E(T)=G(T)-G_S(T),
\end{align}
by measuring the $T$ dependence of $G$ at a constant volume.
Because $G$ is a material constant, it is better to use $G$ rather than $\sigma$ to discuss entropy and energy elasticities.
However, many studies on rubber elasticity have used $\sigma$ because of the difficulty in accurately measuring $G$.\\

\begin{figure}[t!]
 \centering
\includegraphics[width=\linewidth]{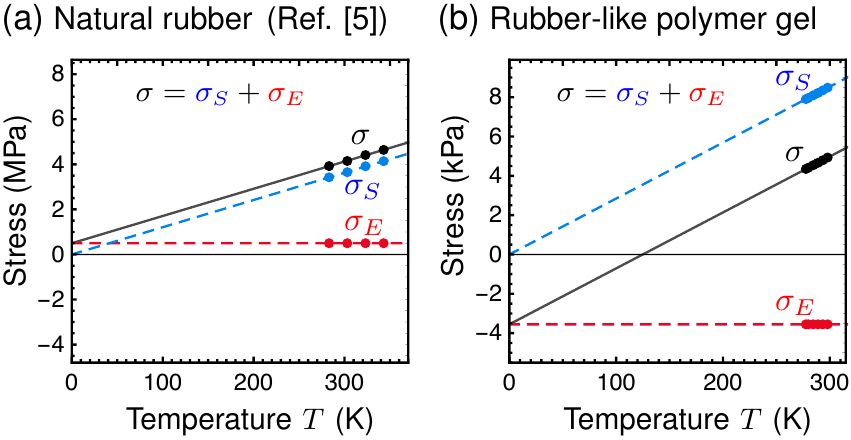}
 \caption{
Irrelevant energy elasticity in natural rubber and relevant negative energy elasticity in a rubberlike polymer gel.\\
(a) Temperature ($T$) dependence of the tensile stress $\sigma$ (black symbols) of vulcanized rubber through stretching measurements under 60\% strain.
The data are taken from Ref.~\cite{Anthony1942}. 
The gray solid line is obtained from a least-squares fit and extrapolated to $T=0$ K.
According to Eq.~(\ref{eq:sigma-vantHoff}), we have
the entropy contribution $\sigma_{S}$ (blue dashed line) and the energy contribution $\sigma_{E}$ (red dashed line), which corresponds to the intercept of the gray solid line.
In the measured temperature range (black symbols), the ratio of $\sigma_{E}$ to $\sigma$ is less than 15\%. 
Similarly, small energy contributions to elasticity were observed in many rubber materials \cite{Meyer1935, Meyer1946, Mark1965, Allen1970, Chen1973}.\\
(b) Typical result of the $T$ dependence of the shear stress $\sigma$ (black symbols) of a polymer gel through rheological measurements under 60\% shear strain $\gamma$.
The gel sample is synthesized by equal-weight mixing of the two kinds of precursors whose molar mass $M$ and concentration $c$ are $20$ kg$/$mol and $60$ g$/$L, respectively.
The gray solid, blue dashed, and red dashed lines are obtained in the same way as in (a).
Notably, we find that $\sigma_{E}$ is a significant negative value.
}
\label{fig:comparison}
\end{figure}

Figure~\ref{fig:comparison} demonstrates how to determine $\sigma_S$ and $\sigma_E$ from experimental data [$\sigma (T)$] for natural rubber [Fig.~\ref{fig:comparison}(a)] and rubberlike polymer gel [Fig.~\ref{fig:comparison}(b)].
Both are highly stretchable as a result of network structures formed by chemically cross-linked polymer chains.
As shown in Fig.~\ref{fig:comparison}(a), it is confirmed that $\sigma\simeq \sigma_S$, in the case of natural and synthetic rubbers, 
by numerous experimental \cite{Meyer1935, Anthony1942, Meyer1946, Mark1965, Allen1970, Chen1973} and theoretical \cite{Duering1994, Everaers1995} studies; $\left|\sigma_E\right|$ is less than a quarter of $\sigma$.
Thus, the elasticity of rubberlike (i.e., hyperelastic) polymer materials has been widely considered to be described primarily as entropy elasticity \cite{Flory1953, Rubinstein2003}.
For example, in polymer gels containing a large amount of solvent, $\sigma \simeq \sigma_S$ has also been postulated \cite{Mark1977, Patel1992, Akagi2013, Zhong2016}; nevertheless, no experimental verification has been reported.\\

In this study, to examine this conventional postulation ($\sigma \simeq \sigma_S$), we measure the temperature dependence of the shear stress and the shear modulus in a rubberlike polymer gel in the as-prepared state.
Here, the gel is a chemical gel (i.e., a covalently bonded gel), and thus the cross-linked structure does not change.
Remarkably, we find that $\sigma_E$ can have a magnitude as large as $\sigma$ but is negative; i.e., $\left|\sigma_E\right| \sim \sigma$ [Fig.~\ref{fig:comparison}(b)].
This means that $G_E$ can have a negative value as large as $G$, although the shear modulus of stable materials is generally bound to be positive.
We further argue that $G_E$ is governed by a vanishing temperature, which is a universal function of the normalized polymer concentration, and $G_E$ vanishes when the solvent is removed.
Our findings would stimulate a re-examination of a vast amount of research on gel elasticity.\\

This paper is organized as follows.
In Sec.~\ref{sec:Materials}, we explain the materials and methods.
In Sec.~\ref{sec:Property}, we present the basic properties of gel elasticity.
In Sec.~\ref{sec:Universal}, we analyze our experimental results and validate the existence of a universal function that governs the energy contribution of gel elasticity.
In Secs.~\ref{sec:Introduction}--\ref{sec:Universal}, we analyze the experimental results based on the general formulas of thermodynamics of deformation under the constant-volume condition, without considering unnecessary microscopic (molecular) assumptions.
In Sec.~\ref{sec:Origin}, we propose a possible microscopic interpretation of the negative energy elasticity.
In Sec.~\ref{sec:Conclusion}, we summarize the main results of this paper.
Several details are described in the appendixes to avoid digressing from the main subject.

\begin{figure}[t!]
 \centering
\includegraphics[width=8.5cm]{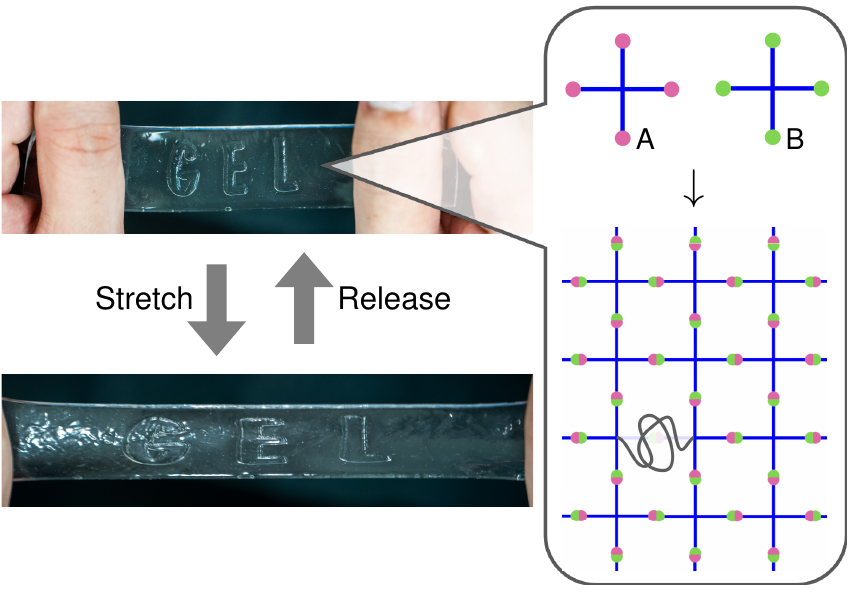}
\caption{
Highly stretchable rubberlike hydrogel with a homogeneous model network (tetra-PEG gel).
This gel is synthesized by AB-type cross-end coupling of two kinds of precursors of the same size.
These precursors are tetra-armed poly(ethylene glycol) (PEG) chains whose terminal functional groups (A and B) are mutually reactive.}
\label{fig:tetra-peg}
\end{figure}

\section{Materials and methods}
\label{sec:Materials}

\subsection{Fabrication of tetra-PEG gels}
\label{sec:Fabrication}

For a model rubberlike polymer gel, we used a tetra-arm poly(ethylene glycol) (tetra-PEG) gel \cite{Sakai2008}, which is highly stretchable and has a homogeneous network structure \cite{Matsunaga2009}.
As shown in Fig.~\ref{fig:tetra-peg}, tetra-PEG gel was synthesized by AB-type cross-end coupling of two kinds of precursors: tetra-arm polymer-chain units of equal size.
For the precursors, we used tetra-maleimide-terminated PEG (tetra-PEG-MA) and tetra-thiol-terminated PEG (tetra-PEG-SH) with molar masses of $M=10$, $20$ (NOF Co., Japan), and $40$ (XIAMEN SINOPEG BIOTECH Co., Ltd., China) kg$/$mol.
We dissolved each precursor (tetra-PEG-MA and tetra-PEG-SH) in phosphate-citrate buffer with pH $3.8$ and a molar concentration of $200$ mM.
Here, the experimental results of the shear modulus $G$ did not depend on the molar concentration and pH.
We obtained gel samples by mixing these precursor solutions (tetra-PEG-MA and tetra-PEG-SH solutions) with equal molar mass $M$ and precursor concentration $c$.
To control the connectivity $p$ after the completion of the chemical reaction,
we nonstoichiometrically tuned the mixing fraction $q_\mathrm{w}$ defined by the weight fraction of tetra-PEG-MA to all precursors (see Appendix~\ref{App:Connectivity}).
We set $c = 30$, $60$, $90$, $120$ g$/$L and $q_\mathrm{w} = 0.50$, $0.55$, $0.60$, $0.65$. 
Here, we define the concentration $c$ as the precursor weight divided by the solvent volume, rather than by the solution volume.
(The reason for adopting this definition is described in Sec.~S1 in the Supplemental Material of Ref.~\cite{Yasuda2020}).
All gel samples used in this study have a homogeneous network structure (thus, the polymer and solvent are completely miscible), as shown in scattering experiments \cite{Matsunaga2009}.

\subsection{Measurement of shear modulus}
\label{sec:Measurement}

\begin{figure}[t!]
 \centering
\includegraphics[width=\linewidth]{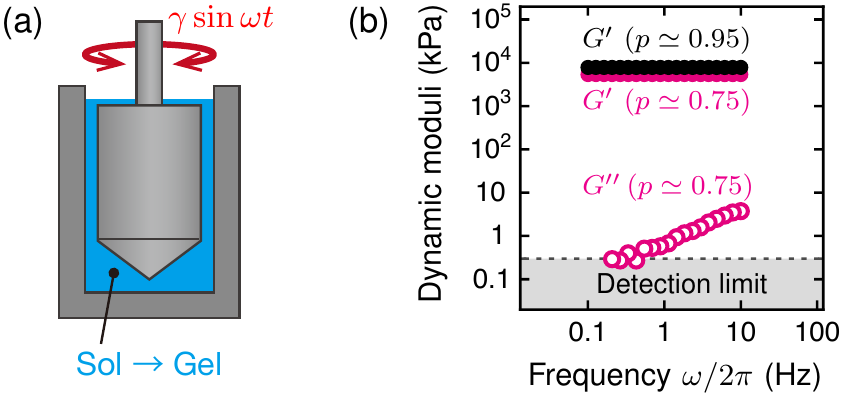}
 \caption{Measurement of shear modulus.\\
(a) Schematic illustration of the double-cylinder dynamic shear rheometer.
Inside the rheometer, a solution of the precursors reacts to form a polymer gel.
After the completion of the chemical reaction, the polymer gel is sheared cyclically in the gap between the inner cylindrical jig and the outer cup.\\
(b) Frequency ($\omega/2\pi$) dependence of the storage modulus $G'$ and loss modulus $G''$.
The molar mass $M$ and concentration $c$ of the precursors are $20$ kg$/$mol and $60$ g$/$L, respectively.
The greater the connectivity $p$ is, the lower $\tan\delta$ is.
($p$ is defined in the main text).
For example, when $p\simeq0.95$ (a nearly perfect network structure), $G''$ is underdetected.}
\label{fig:measurement}
\end{figure}

We measured the shear modulus $G$ using a dynamic shear rheometer (MCR301 and MCR302, Anton Paar, Austria), as shown in Fig.~\ref{fig:measurement}(a).
Immediately after mixing the two kinds of precursor solutions, we poured the resulting solution into the gap within the double cylinder of the dynamic shear rheometer.
Then, we measured the time courses of the storage modulus $G'$ and loss modulus $G''$ at $298$ K with the applied strain ($\gamma$) of $1$\%.
After $G'$ reached equilibrium, indicating the completion of the chemical reaction between maleimide and thiol, we measured the temperature ($T$) dependence of $G'$ from $T=298$ to $278$ K and then from $278$ to $298$ K.
There was almost no hysteresis for the mentioned temperature range.
In all samples, the obtained $G'$ is independent of the frequency ($\omega/2\pi$) below $10$ Hz, and the loss tangent $\tan\delta =G''/G'$ is at most $10^{-2}$ at $1$ Hz, as shown in Fig.~\ref{fig:measurement}(b).
Thus, we regard $G'$ at $1$ Hz as the (equilibrium) shear modulus $G$ given by $G=\lim_{\omega\to0}G'(\omega)$.\\

The aforementioned rheological measurement allows us to satisfy the constant-volume condition
necessary to determine the entropy and energy contributions ($\sigma_{S}$, $\sigma_{E}$, $G_{S}$, and $G_{E}$).
The factors that contribute to volume changes are shear deformation and temperature change.
The shear deformation can suppress the volume change caused by the decrease in internal pressure \cite{Flory1953,Meyer1946}.
Setting $278\, \mathrm{K} \leq T\leq 298\, \mathrm{K}$ guarantees a negligible volume change for the following three reasons:
(i) If $T<273$ K, the water freezes; 
(ii) if $T$ considerably exceeds $298$ K, the gel shrinks owing to the large elastic contribution to the osmotic pressure \cite{Tanaka1978};
(iii) the gel hardly thermally expands because it contains a large amount of water as a solvent.
A detailed quantitative analysis is given in Appendix~\ref{App:Volume}.
The relative volume change due to temperature change within $278\, \mathrm{K} \leq T\leq 298\, \mathrm{K}$ is 
of the order of $10^{-3}$, and the relative volume change due to shear deformation is negligible in comparison.

\begin{figure}[t!]
 \centering
\includegraphics[width=8cm]{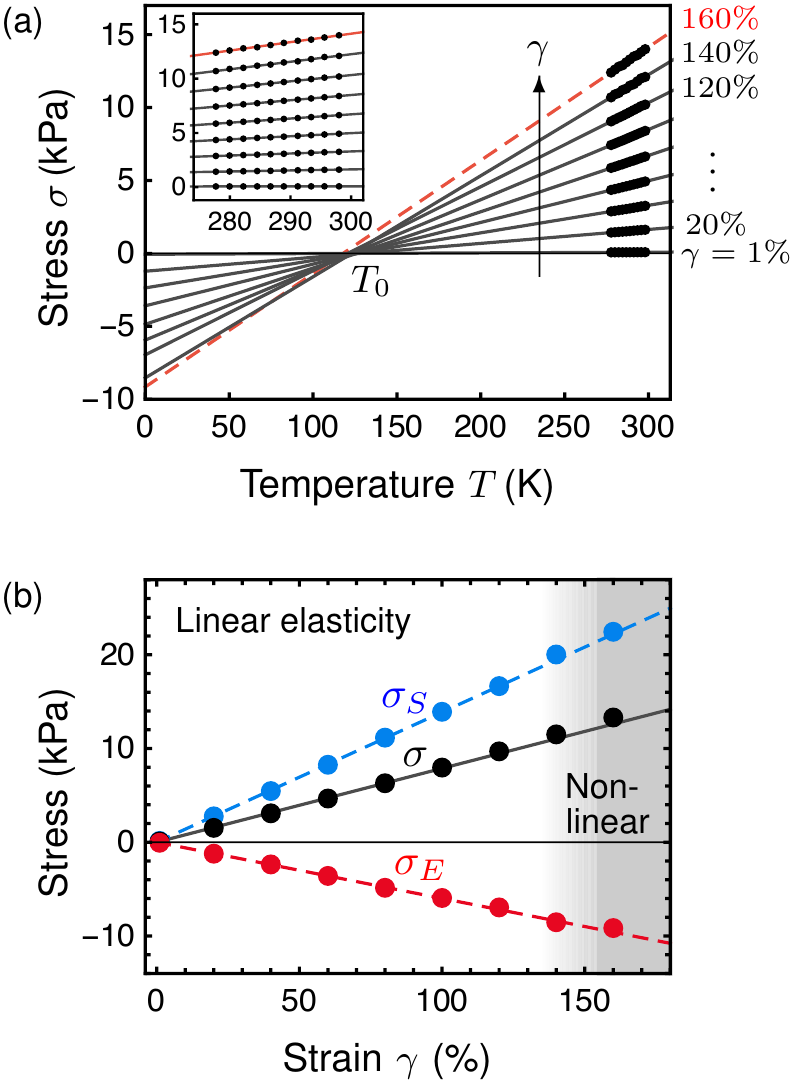}
 \caption{
Polymer gel exhibiting a wide range of linear elasticity under shear deformation.\\
(a) Temperature ($T$) dependence of (shear) stress $\sigma$ under fixed (shear) strain $\gamma$ ($1, 20, 40, 60, 80, 100, 120, 140, 160$, and $180\%$) within $278\, \mathrm{K} \leq T\leq 298\, \mathrm{K}$.
The data for $\gamma=60\%$ correspond to Fig.~\ref{fig:comparison}(b).
The straight lines are obtained from the least-squares fit and extrapolated to $T=0$ K.
Within the measured range, $\sigma$ is a nearly linear function of $T$ at each $\gamma$ (inset). 
The gray solid extrapolation lines ($\gamma\leq 140\%$) pass through the vanishing temperature $T_0$ on the $T$ axis.\\
(b) Stress ($\sigma$)--strain ($\gamma$) curve (black symbols) with $\sigma_{S}$ (blue symbols) and $\sigma_{E}$ (red symbols).
The data are extracted from (a) at $T=288$ K.
The $\sigma$, $\sigma_{S}$, and $\sigma_{E}$ show linear elasticity over a wide range up to $\gamma =140\%$.
The linearity of $\sigma_{S}$ and $\sigma_{E}$ to $\gamma$ corresponds to the $\gamma$ independence of $T_0$.
}
\label{fig:linear}
\end{figure}

\section{Basic properties of gel elasticity}
\label{sec:Property}

Figure~\ref{fig:linear}(a) shows the $T$ dependence of $\sigma$ under fixed strain $\gamma$ for a polymer gel and demonstrates the following two facts.
First, $\sigma$ is a nearly linear function of $T$ within the measured range.
Thus, we regard the slope of the linear fitting of the $\sigma$-$T$ relation as $\partial \sigma/\partial T$ in Eq.~(\ref{eq:sigma-vantHoff}).
Second, all linear extrapolations of $\sigma=\sigma (T)$ for $\gamma \leq 140\%$ (gray solid lines) pass through a common point ($T_0$) on the $T$ axis.
We call this proper temperature $T_0$ the ``vanishing temperature''.
We emphasize that the ``actual'' stress $\sigma$ does not follow the extrapolation lines (gray solid lines) at low temperatures away from the measured temperature ($< 278$ K), and it certainly does not vanish at $T_0$. 
We extrapolate the $\sigma$-$T$ relation only to calculate the energy and entropy contributions of $\sigma$.\\

The polymer gel (tetra-PEG gel) is an ideal rubberlike (i.e., hyperelastic) material in the sense that the stress-strain relation [the black symbols in Fig.~\ref{fig:linear}(b)] exhibits a wide range of linear elasticity ($\gamma \lesssim 140\%$).
Thus, over a wide range, the shear modulus $G$ [defined in Eq.~(\ref{eq:def:G})] describes the elasticity of the polymer gel.
This ideal linear elasticity of the polymer gel implies that the volume is indeed constant under shear deformation and is advantageous for investigating elasticity compared to natural and synthetic rubbers (the relative volume change is evaluated in Appendix~\ref{App:Volume}).\\

Figure~\ref{fig:linear}(b) demonstrates that the polymer gel exhibits negative energy elasticity; the energy contribution $\sigma_E$ of a polymer gel can be a significant negative value, whose absolute value is comparable to $\sigma$.
Here, $\sigma_S$ (blue symbols) and $\sigma_E$ (red symbols) are calculated from $\sigma (T)$ in Fig.~\ref{fig:linear}(a) using Eq.~(\ref{eq:sigma-vantHoff}).
Figure~\ref{fig:linear}(b) also demonstrates that $\sigma_S$ and $\sigma_E$ are linear with respect to $\gamma$ for a wide range of the strain ($\gamma \lesssim 140\%$).
Thus, over a wide range, the entropy and energy contributions to the shear modulus 
[$G_{S}$ and $G_{E}$ defined in Eqs.~(\ref{eq:def:GS}) and (\ref{eq:def:GE}), respectively] describe the entropy and energy elasticities, respectively.
We note that the negative energy elasticity is observed in the gel with a homogeneous network structure, but not under special circumstances such as phase separation (see Sec.~\ref{sec:Fabrication}).

\begin{figure*}[t!]
\centering
\includegraphics[width=17.58cm]{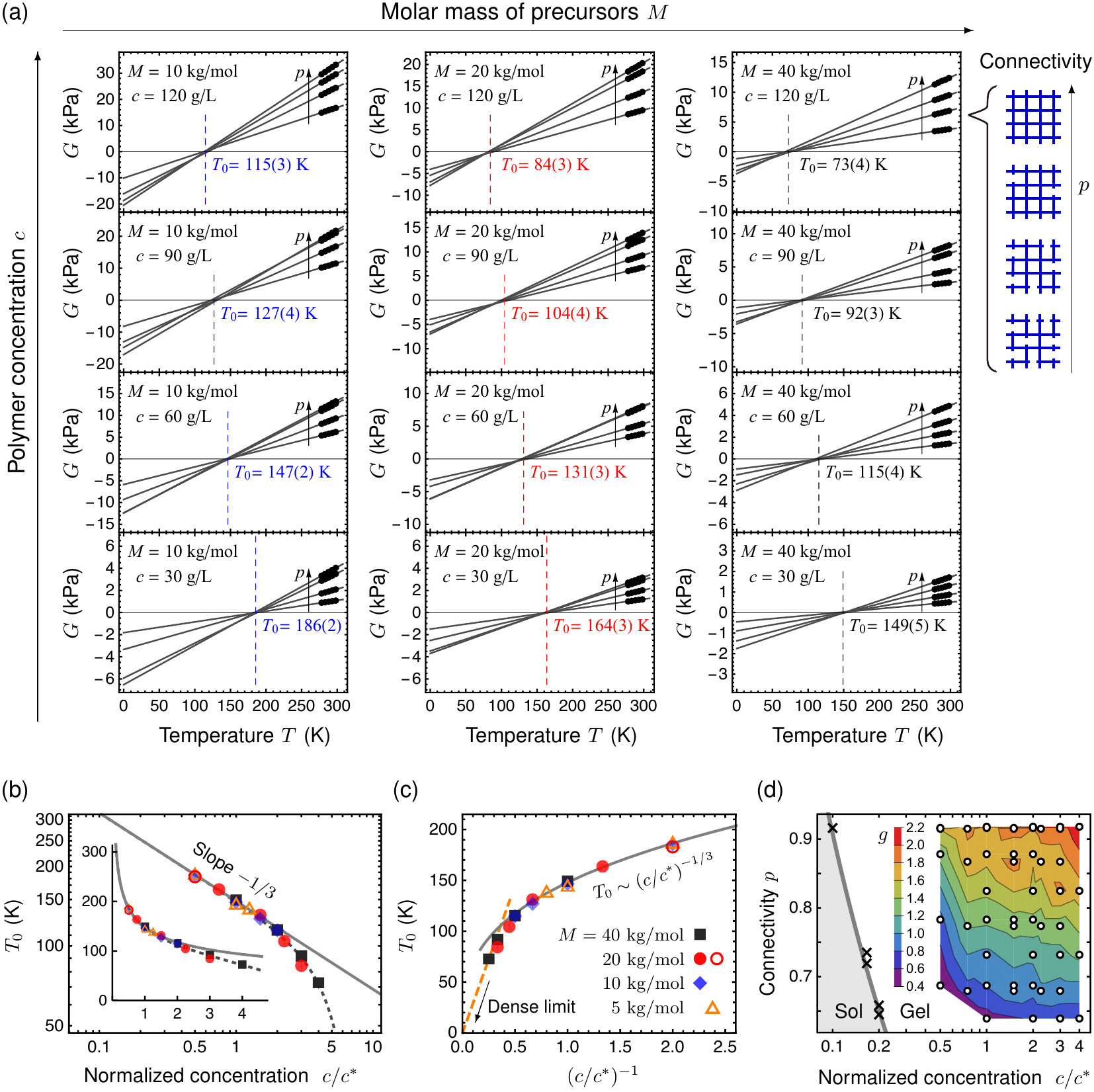}
\caption{
The existence of a universal function that governs the energy contribution of gel elasticity.\\
(a) Temperature ($T$) dependence of shear modulus $G$.
We obtain each gray line from a least-squares fit of each sample, which is characterized by the three parameters of the precursors:
the molar mass $M$, the concentration $c$, and the connectivity $p$.
All gray lines that have the same $M$ and $c$ pass through a vanishing temperature $T_0$ on the $T$ axis, 
which leads to Eq.~(\ref{eq:Gbya}).
The value of $T_0$ in each graph is the average of the four samples with different values of $p$, and the values in parentheses represent the standard deviation.\\
(b, c) Normalized concentration ($c/c^{*}$) dependence of $T_0$.
We set $c^{*}= 80$, $60$, $40$, and $30$ g/L for $M=5$, $10$, $20$, and $40$ kg/mol, respectively, to construct the master curve.
The orange triangles, blue diamonds, red circles, and black squares represent $M=5$, $10$, $20$, and $40$ kg/mol, respectively.
Each filled symbol represents the average of four samples taken from (a).
Additionally, each open symbol represents the value of one equal-weight mixing sample ($p\simeq 1$).
(b) Log-log (main panel) and linear (inset) plots of $c/c^{*}$ vs $T_0$, which demonstrate the scaling law $T_0\sim (c/c^{*})^{-1/3}$ in the dilute regime ($c/c^{*}<1$).
(c) A linear plot of the inverse normalized concentration $(c/c^{*})^{-1}$ vs $T_0$, which demonstrates that $T_0$ becomes nearly zero in the dense limit [$(c/c^{*})^{-1} \to 0$].
This result agrees with $|G_{E}|\ll G_{S}$ for rubber elasticity as given in previous studies \cite{Meyer1935, Anthony1942, Meyer1946, Mark1965, Chen1973}.\\
(d) Connectivity $p$ and $c/c^{*}$ dependence of $g$.
We obtain the contour plot from data points (white circles) that represent samples shown in (a).
The sol-gel transition line (gray thick line) is the interpolation of data (black crosses) taken from Ref.~\cite{Hayashi2017}.
}
\label{fig:universal}
\end{figure*}

\section{Universal function governing energy elasticity}
\label{sec:Universal}

\subsection{Vanishing temperature}
To investigate the relationship between the negative energy elasticity of the polymer gel and the microscopic structure of the polymer network,
we independently and systematically control three parameters of the precursors: the molar mass $M$, the concentration $c$, and the connectivity $p$ (see Sec.~\ref{sec:Fabrication}). 
Here, we define $p$ ($0\leq p\leq1$) as the fraction of the reacted terminal functional groups to all the terminal functional groups and control $p$ by nonstoichiometrically mixing two mutually reactive precursor solutions (see Appendix~\ref{App:Connectivity}).
In the polymer network after the completion of the chemical reaction, $c$ and $M$ correspond to the polymer (network) concentration and double the molecular weight between crosslinks, respectively.\\ 

From the experimental results shown in Fig.~\ref{fig:universal}(a), we find two features: 
(i) $G$ is a nearly linear function of $T$ in the measured range, 
and (ii) the vanishing temperature $T_0$ is independent of $p$.
These features indicate that
\begin{equation}
G(T,M,c,p)=a(M,c,p) \Bigl[ T-T_0(M, c)\Bigr],
\label{eq:Gbya}
\end{equation}
where we introduce a prefactor $a=a(M, c, p)$.
According to Eq.~(\ref{eq:Gbya}), in the measured range, the entropy contribution
\begin{equation}
G_{S}(T,M,c,p)=a(M,c,p) T
\end{equation}
is a linear function of $T$, and the energy contribution
\begin{equation}
G_{E}(T,M,c,p)=-a(M,c,p) T_{0}(M, c)
\end{equation}
is independent of $T$ and governed by $T_0$.\\

By analyzing the systematic results shown in Fig.~\ref{fig:universal}(a), we reveal a law governing $T_0$. 
Figure~\ref{fig:universal}(b) demonstrates that all the results at different values of $M$ and $c$ collapse onto a single master curve, indicating
\begin{equation}
T_0(M,c)=T_0\left(\frac{c}{c^*}\right).
\label{eq:T0}
\end{equation}
Here, $c^*=c^*(M)$ is the normalization factor chosen to construct the master curve.
It is notable that $c^*(M)$ is 

\newpage
\quad\newpage
\quad\newpage

\noindent
in close agreement with the overlap concentration of the precursors $c_\mathrm{vis}^*(M)$ obtained by the viscosity measurement \cite{Akagi2013}.
Here, the overlap parameter $c/c_\mathrm{vis}^*$ determines the dilute ($c/c_\mathrm{vis}^*<1$) and semidilute ($c/c_\mathrm{vis}^*>1$) regimes of polymer concentration \cite{Rubinstein2003}. 
This agreement and Eq.~(\ref{eq:T0}) invoke the osmotic pressure in the polymer solution, 
which is represented by a universal function of $c/c_\mathrm{vis}^*$ in the dilute ($c/c_\mathrm{vis}^*<1$) and semidilute ($c/c_\mathrm{vis}^*>1$) regimes \cite{Cloizeaux1975}.
We note that the maximum polymer volume fraction in this study is approximately $0.1$, and it is unclear whether Eq.~(\ref{eq:T0}) holds at higher polymer concentrations ($c/c_\mathrm{vis}^*\gg 1$).\\

We consider the dilute and dense regimes.
In the dilute regime ($c/c^{*}<1$), we find a scaling law 
\begin{equation}
T_0\sim \left(\frac{c}{c^{*}}\right)^{-1/3},
\label{eq:T0Dilute}
\end{equation}
as shown in Fig.~\ref{fig:universal}(b).
Because $c^{-1/3}$ seems to be proportional to the linear distance between crosslinks $l$, we have $T_0\sim l$.
This fact is important in conjecturing the microscopic (molecular) interpretation of the negative energy elasticity (see Sec.~\ref{sec:Origin}).
If $T_0$ follows the scaling law in Eq.~(\ref{eq:T0Dilute}) below the measured $c/c^{*}$ range, $T_{0}$ reaches the measured temperature ($T \gtrsim 280$~K) at $c/c^{*}\simeq 0.12$. 
Thus, if $c<0.12c^{*}$, tetra-PEG gels are expected to be mechanically unstable because $G<0$.
This expectation is consistent with a previous study that reported that tetra-PEG gels cannot be formed below $c/c^{*}\simeq 0.1$ around $300$ K \cite{Hayashi2017}.\\

In the dense regime ($c/c^{*} \gg 1$), Fig.~\ref{fig:universal}(c) shows that as $(c/c^{*})^{-1} \to 0$, which means that the solvent is removed, $T_0$ decreases, approaching nearly zero.
This result is in agreement with previous studies on natural and synthetic rubbers without solvent; the absolute value of the energy contribution ($aT_0$) is much smaller than the value of the entropy contribution ($aT$) \cite{Meyer1935, Anthony1942, Meyer1946, Mark1965, Allen1970, Chen1973}. 
In other words, this result suggests that the presence of solvent is the origin of energy elasticity in the polymer gels, as discussed in Sec.~\ref{sec:Origin}.

\subsection{Prefactor of shear modulus}

Using dimensional analysis, we determine the functional form of $a=a(p, M, c)$.
Since $a$ has the same dimension as $cR/M$, 
the dimensionless combination is $g \equiv aM/(cR)$,
where $R$ is the gas constant.
Then, $g$ depends on the dimensionless parameters composed of $p$, $M$, and $c$; that is, $g=g(p, c/c^{*})$.
Figure~\ref{fig:universal}(d) validates this dimensional analysis.
Substituting $g=g(p, c/c^{*})$ and Eq.~(\ref{eq:T0}) into Eq.~(\ref{eq:Gbya}), we have 
\begin{equation}
G(T,M,c,p)=\frac{cR}{M}g\left(p, \frac{c}{c^{*}}\right)\left[T-T_0\left(\frac{c}{c^{*}}\right)\right].
\label{eq:main-result}
\end{equation}
Although $g(p, c/c^{*})$ and $T_ {0}(c/c^{*})$ depend on the kinds of polymer chains and solvents that constitute the polymer gels, Eq.~(\ref{eq:main-result}) generally represents $G$ in homogeneous polymer gels.
For example, Fig.~\ref{fig:universal}(c) gives $T_ {0}(c/c^{*})$ for the tetra-PEG gel.
We note that Eq.~(\ref{eq:main-result}) holds for the narrow temperature range (such as $278\, \mathrm{K} \leq T\leq 298\, \mathrm{K}$), where the polymer gels are in a rubbery state and the relative volume change due to temperature change is negligible (see Appendix~\ref{App:Volume}).\\

The behavior of the contour lines of $g(p,c/c^{*})$ in Fig.~\ref{fig:universal}(d) is consistent with recent experiments on the polymer gels for the dilute ($c/c^{*}<1$) and semidilute ($c/c^{*}>1$) regimes.
In the dilute regime ($c/c^{*}<1$), Fig.~\ref{fig:universal}(d) shows that the contour lines of $g(p,c/c^{*})$ are nearly parallel to the sol-gel transition line~\cite{Hayashi2017} corresponding to $g(p,c/c^{*})=0$.
Thus, the contour lines are consistent with previous experiments~\cite{Sakai2016,Hayashi2017}.
There seems to be no theory (e.g., percolation and mean-field theories) that quantitatively explains the dependence of $g$ on $c/c^{*}$~\cite{Sakai2016}.
The behavior of $g(p,c/c^{*})$ is qualitatively considered to be caused by elastically ineffective connections such as intramolecular bonds and loops \cite{Sakai2016, Zhong2016, Yoshikawa2019}.\\

In the semidilute regime ($c/c^{*}>1$),
Fig.~\ref{fig:universal}(d) shows that $g$ is almost independent of $c/c^{*}$, and 
we find that $g(p, c/c^{*})\simeq 2.4\,\xi(p)$ from the experimental data (see Appendix~\ref{App:Bethe-Approximation}).
Here, the number per precursor of elastically effective cycles $\xi(p)$ satisfies $0\leq\xi(p)\leq1$ and is calculated using the Bethe approximation \cite{Macosko1976,Miller1976}.
Equation~(\ref{eq:main-result}) with $g(p, c/c^{*})\simeq 2.4\,\xi(p)$ leads to $G(p) \sim \xi(p)$,
which is consistent with the previous experiments \cite{Nishi2017,Yoshikawa2019}.
We remark that the entropy contribution obtained in our study,
$G_{S}=cRg(p)T/M$,
is $2.4$ times as large as $G_{S}^{\mathrm{phan}}=cR\xi(p) T/M$,
which is predicted by the phantom network model \cite{James1953,Rubinstein2003}.

\begin{figure}[t!]
\centering
\includegraphics[width=8cm]{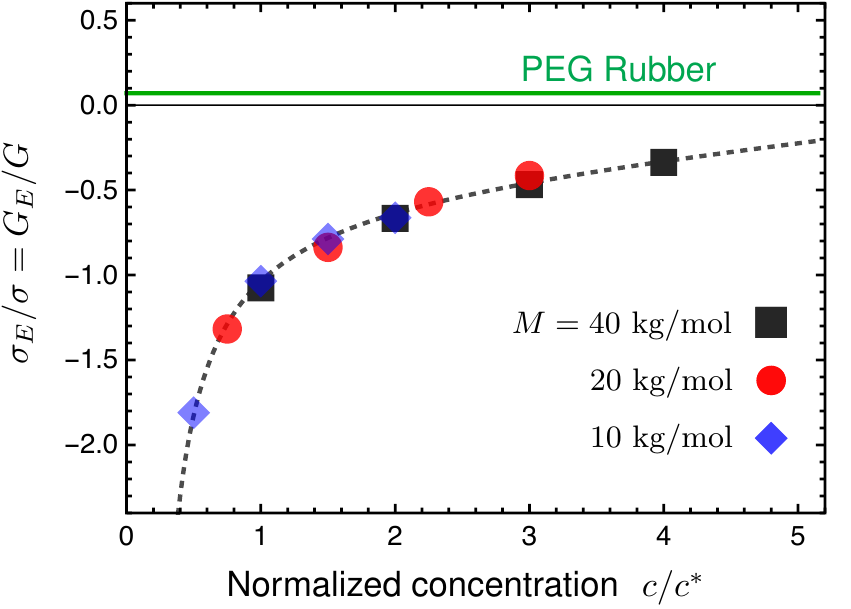}
\caption{
Strong dependence of the energy contribution for polymer gel elasticity ($\sigma_{E}/\sigma=G_E/G$) on the normalized polymer concentration $c/c^{*}$.
The data are taken from Fig.~\ref{fig:universal}(a), and the symbols are the same as those in Figs.~\ref{fig:universal}(b) and \ref{fig:universal}(c).
The energy contribution ($\sigma_{E}/\sigma$) of the polymer gel is considerably different from that of poly(ethylene glycol) (PEG) rubber ($\sigma_{E}/\sigma=0.07\pm0.01$ \cite{Mark1965}, green line).
This discrepancy suggests that the explanation of energy contribution in previous studies of rubber elasticity [e.g., the rotational isomeric state (RIS) model \cite{Flory1969} as shown in Fig.~\ref{fig:origin}(a)] is not applicable to that in this study of gel elasticity.
On the other hand, $\sigma_{E}/\sigma$ tends to approach that of PEG rubber in the dense limit ($c/c^{*} \to \infty$).
}
\label{fig:energy}
\end{figure}

\begin{figure}[t!]
\centering
\includegraphics[width=\linewidth]{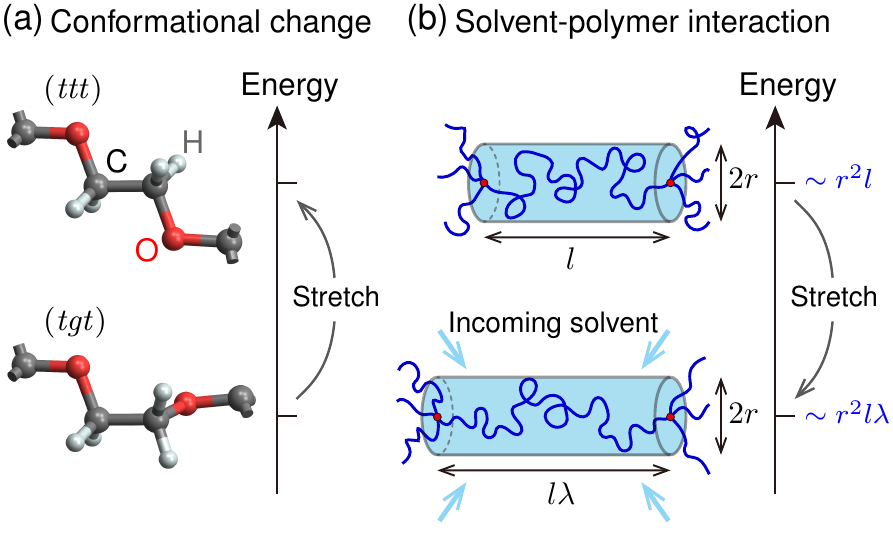}
\caption{
Possible scenarios for the microscopic origin of negative energy elasticity.\\
(a) Energy change accompanying the conformational change with deformation of a PEG chain,
which has been considered to be the origin of ``positive'' energy elasticity for PEG rubber [the green line in (a)] \cite{Mark1965}.
When the PEG chain is stretched, the \textit{trans-gauche-trans} (\textit{tgt}) conformation around successive O--C--C--O bonds is transformed to the \textit{trans-trans-trans} (\textit{ttt}) conformation.
Because the conformational energy of \textit{trans} around the C--C bond is higher than that of \textit{gauche} originating from the dispersion interaction between the adjoining oxygen atoms \cite{Mark1965}, the energy elasticity is positive.\\
(b) Energy change accompanying the attractive solvent-polymer interaction with deformation of a polymer (PEG) chain between crosslinks, causing ``negative'' energy elasticity in the polymer gel.
In the dilute regime ($c/c^{*}<1$), the ``territory'' of a chain (blue curves) between crosslinks (red points) can be roughly described as cylindrical in shape (light blue region) with radius $r$ and height $l$.
When the chain is mechanically stretched at a stretch ratio $\lambda$, the number of solvent molecules interacting with the chain increases, decreasing the total energy of the solvent-polymer interaction.
}
\label{fig:origin}
\end{figure}

\section{Microscopic origin of negative energy elasticity}
\label{sec:Origin}

In this section, we discuss the microscopic (molecular) origin of negative energy elasticity.
A remarkable feature of negative energy elasticity in polymer gels, discovered in this study, is its strong concentration ($c$) dependence of the energy contribution ($\sigma_{E}/\sigma=G_E/G$), as shown in Fig.~\ref{fig:energy}.
This concentration dependence is expected to be more difficult to observe in gels with large polymer volume fractions.
In fact, although some studies \cite{Mark1976, Allen1970} measured the energy elasticity in swollen rubbers with a polymer volume fraction of at least $0.2$, they did not observe the concentration dependence of the energy contribution.
On the other hand, in this study, we measure the energy elasticity in the polymer gels with a volume fraction of less than or around $0.1$, and we find a strong concentration dependence.\\

Figure~\ref{fig:energy} reveals that the conformational energy change, which is
 considered to be the origin of the energy elasticity in rubber, cannot explain the microscopic origin of negative energy elasticity in a rubberlike gel.
The complete evaporation of water from the tetra-PEG gel (used in this study) corresponds to PEG rubber, which has a small, positive energy contribution ($\sigma_{E}/\sigma=0.07\pm0.01$, green line) \cite{Mark1965}.
This energy contribution is interpreted to be due to the conformational energy change around the C--C bond [Fig.~\ref{fig:origin}(a)] \cite{Mark1965, Chen1973, Mark1976} through the rotational isomeric state (RIS) model \cite{Flory1969} (further details are given in Appendix~\ref{App:History-rubber}).
However, the small and positive energy contribution is considerably different from our results, as shown in Fig.~\ref{fig:energy}; $\sigma_{E}/\sigma$ for the tetra-PEG gel is large and negative ($\sigma_{E}/\sigma=G_E/G=-1.81\pm 0.04$ at the maximum) and has a strong dependence on $c$.
Thus, we suggest that the internal energy change with deformation originates mainly from some kind of intermolecular (that is, polymer-polymer, solvent-polymer, and solvent-solvent) interaction.
In addition, since the volume is constant (the components remain constant), the Flory-Huggins theory (and the $\chi$ parameter) \cite{Flory1953,deGennes1979,Rubinstein2003} cannot explain the interaction energy change with deformation (see Appendix~\ref{App:Thermodynamics}).
Therefore, we need a new microscopic interpretation of negative energy elasticity.\\

We propose a possible microscopic interpretation of the negative energy elasticity in a rubberlike gel [Fig.~\ref{fig:origin}(b)];
the negative $G_{E}/G$ value in the gel and the scaling law in Eq.~(\ref{eq:T0Dilute}) originate from the fact that the stretching of a polymer chain increases the number of solvent molecules that interact attractively with polymer chains.
The top of Fig.~\ref{fig:origin}(b) shows a single polymer chain (blue curve) between crosslinks (red points) in the dilute regime ($c/c^{*}<1$), where most of the chains are isolated.
We define the “territory” of a polymer chain as the region containing solvent molecules that interact with the chain and assume that the territory is roughly described as cylindrical in shape (light blue region) with radius $r$ and height $l$.
As shown at the bottom of Fig.~\ref{fig:origin}(b), when the polymer chain is mechanically stretched at a stretch ratio $\lambda$, the territory increases; that is, the number of solvent molecules interacting with the chain increases.
Here, the solvent molecules come from outside the original (i.e., undeformed) territory.
Under small deformation, the radius $r$ remains nearly constant because the length of a chain in the direction perpendicular to the stretching direction does not change \cite{deGennes1979, Rubinstein2003}.
Assuming that the total energy of the solvent-polymer interaction per chain is proportional to the volume of the territory, the solvent-polymer interaction energies in the undeformed and deformed chains are proportional to $r^2l$ and $r^2l\lambda$, respectively.
Thus, when a single polymer chain is stretched, the change in the total energy of the solvent-polymer interaction is proportional to $r^2l(\lambda -1)$.
Therefore, the energy elasticity of the gel ($G_E\sim T_{0}$) is proportional to the linear distance between the crosslinks $l$, i.e., $T_{0}\sim l$.
Supposing $l\sim c^{-1/3}$, we have $T_{0}\sim c^{-1/3}$, which is consistent with the scaling law in Eq.~(\ref{eq:T0Dilute}) for a fixed $M$ [and a fixed $c^*(M)$].\\

The proposed microscopic interpretation is also consistent with the experimental results in semidilute and dense regimes.
In the semidilute regime ($c/c^{*}>1$), the territories [light blue regions in Fig.~\ref{fig:origin}(b)] overlap with each other.
Thus, the dependence of $T_{0}$ on $c$ falls below the scaling law ($T_{0}\sim c^{-1/3}$) as shown in Fig.~\ref{fig:universal}(b).
In the dense regime ($c/c^{*}\gg 1$), the solvent diminishes, and the effect of the solvent-polymer interaction on energy elasticity becomes negligible compared to that of the conformational energy change [Fig.~\ref{fig:origin}(a)], which is consistent with the results shown in Figs.~\ref{fig:universal}(c) and \ref{fig:energy}.\\

Although the proposed microscopic interpretation appears to explain the experimental results well, further investigations are needed, as we only performed macroscopic measurements in this study.
For example, molecular-scale experiments (such as light scattering and single-chain experiments) and numerical simulations (such as molecular dynamics simulations) will better reveal the microscopic origin of the negative energy elasticity.

\section{Concluding remarks}
\label{sec:Conclusion}

In conclusion, we have discovered that the energy contribution to shear modulus ($G_E$), which is negligible in rubbers, can be a significant negative value in rubberlike polymer gels containing a large amount of solvents (with the polymer volume fraction of less than or around $0.1$) in the as-prepared state.
Further systematic experiments with various network structures have revealed that the shear modulus $G$ is simply described by Eq.~(\ref{eq:main-result}), and $G_E$ is governed by a vanishing temperature $T_0$, which is a universal function of the normalized polymer concentration $c/c^{*}$.
The vanishing temperature $T_0$ exhibits a scaling law of $T_0 \sim (c/c^{*})^{-1/3}$ in the dilute regime ($c/c^{*}<1$).
Based on this scaling law, we have suggested the microscopic origin of negative energy elasticity: It emerges from the interaction between the polymer chain and the solvent, as shown in Fig.~\ref{fig:origin}(b).
To establish this origin from a microscopic (molecular) description and to verify whether our findings are universal in other polymer gels, further studies are needed.\\

Our findings have essential implications for past research on gel elasticity, which was previously thought to be the same as the rubber elasticity.
For example, in the previous paper using tetra-PEG gels \cite{Akagi2013}, $G \simeq G_S$, i.e., $G_E \simeq 0$, was assumed, and the dependence of the shear modulus $G$ on the polymer concentration $c$ was interpreted as the crossover between the phantom \cite{James1953,Rubinstein2003} and affine \cite{Flory1953} network models.
However, our results [Figs.~\ref{fig:universal}(b) and \ref{fig:universal}(c) and Eq.~(\ref{eq:main-result})] point out that the above assumption is invalid, and ``the crossover'' does not mean the phantom-affine crossover but originates from the dependence of $T_0$ on $c$.
As this example shows, our study provides a new perspective on gel elasticity and urges re-examinations of a vast amount of previous research on gel elasticity.


\begin{acknowledgments}
We thank Kenji~Urayama and Tetsuo~Yamaguchi for their useful comments.
N.S. thanks Shin-ichi~Sasa and Shigeyuki~Komura for their useful comments.
This work was supported by the Japan Society for the Promotion of Science (JSPS) through Grants-in-Aid for 
Early Career Scientists grant number 19K14672 to N.S.,
Scientific Research (B) grant number 18H02027 to T.S.,
and Scientific Research (S) grant number 16H06312 to U.C. 
This work was also supported by the Japan Science and Technology Agency (JST) CREST grant number JPMJCR1992 to T.S. and COI grant number JPMJCE1304 to U.C.
\end{acknowledgments}

\section*{Author contributions}
T.S. planned and supervised the whole project. 
Y.Y. designed and performed the experiments. 
N.S. developed the theoretical framework.
U.C. contributed to discussions throughout the project.
Y.Y. and N.S. analyzed and interpreted the results and wrote the manuscript.

\appendix
\section{Control of the connectivity}
\label{App:Connectivity}

To obtain tetra-PEG gels with different values of connectivity $p$ after the completion of the chemical reaction, 
we nonstoichiometrically mix the two kinds of precursor solutions in weight fractions of tetra-PEG-MA to all precursors $q_\mathrm{w}$.
Here, $p$ ($0\leq p\leq1$) is the fraction of reacted maleimide and thiol groups to all functional groups.
Assuming that almost all the minor terminal groups react, we have \cite{Yoshikawa2019}
\begin{equation}
p=
\begin{cases}
2qp_\mathrm{MA} ~~ &\mathrm{for}~~0\leq q \leq \dfrac{p_\mathrm{SH}}{p_\mathrm{MA}+p_\mathrm{MA}},\\
2(1-q)p_\mathrm{SH} ~~ &\mathrm{for}~~\dfrac{p_\mathrm{SH}}{p_\mathrm{MA}+p_\mathrm{MA}} \leq q \leq 1,
\end{cases}
\label{eq:p}
\end{equation}
where $p_\text{MA}$ and $p_\text{SH}$ are the terminal functionalization fractions of tetra-PEG-MA and tetra-PEG-SH, respectively.
Here, $q$ is the molar fraction of tetra-PEG-MA to all precursors, and it is determined from $q_\mathrm{w}$ by considering the molecular-weight distributions and functionalities of the precursors.
Further details of this method are described in Ref.~\cite{Yoshikawa2019}.

\section{Volume change due to temperature change and shear deformation}
\label{App:Volume}

To demonstrate that the experiment in this study was conducted with a negligible volume change, we evaluate the relative volume change (i.e., the volume strain) due to temperature change and shear deformation.
First, we evaluate the relative volume change of tetra-PEG gels with a change in temperature during free thermal expansion at $1$~atm.
The densities of the tetra-PEG gels are considered to be equal to the densities of the aqueous PEG solution with the same polymer concentration $c$.
According to Ref.~\cite{Gonzalez-Tello1994}, the densities of aqueous PEG solutions at a temperature $T=277$~K [$\rho(277 \mathrm{K})$] and at a temperature $T=298$~K [$\rho(298 \mathrm{K})$] can be calculated from the following formulas, respectively:
\begin{equation}
\begin{aligned}
\rho(277 \mathrm{K}) / (\mathrm{g}/\mathrm{cm}^{3})&=1.0000+0.19820 w,\\
\rho(298 \mathrm{K}) / (\mathrm{g}/\mathrm{cm}^{3})&=0.99707+0.17441 w.
\end{aligned}
\label{eq:density}
\end{equation}
Here, $w=c/(c+997.07 \, (\mathrm{g}/\mathrm{L}))$ is the mass fraction of PEG.
In this study, experiments were conducted in the temperature range of $278$~K to $298$~K. 
However, to make use of literature values, we compare densities in a slightly wider temperature range ($277$~K and $298$~K).
In addition, in this study, $w$ at each sample is independent of $T$ because all the samples are prepared at $T=298$~K.
The relative volume change caused by a decrease in temperature from $298$~K to $278$~K at 1 atm is given as
\begin{equation}
\frac{\Delta V}{V}
\equiv\frac{V(298 \mathrm{K})-V(277 \mathrm{K})}{V(298 \mathrm{K})}
=1-\frac{\rho(298 \mathrm{K})}{\rho(277 \mathrm{K})},
\label{eq:VolumeChange}
\end{equation}
where $V(277 \mathrm{K})$ and $V(298 \mathrm{K})$ are the volumes at $277$~K and $298$~K, respectively.
By using Eqs. (\ref{eq:density}) and (\ref{eq:VolumeChange}), we obtain $\Delta V/V\sim 10^{-3}$, as summarized in Table~\ref{tab1}.
Here, $m\sim n$ means that the quantities $m$ and $n$ have the same order of magnitude.\\

\begin{table}[t]
\caption{Relative volume change ($\Delta V/V$) with change of temperature in free thermal expansion at atmospheric pressure (1 atm), based on Ref.~\cite{Gonzalez-Tello1994}.}

\label{tab1}
\begin{ruledtabular}
\begin{tabular}{ccccc}
$c$ & $w$ & $\rho (277 \mathrm{K})$ & \, $\rho (298 \mathrm{K})$ &  $\Delta V/V$  \\
$(\mathrm{g}/\mathrm{L})$ & $(-)$ & $(\mathrm{g}/\mathrm{cm}^{3})$ & $(\mathrm{g}/\mathrm{cm}^{3})$ &  $(-)$  \\
\hline
  $30$  & $0.0292$ & 1.0058  & 1.0022  & 0.0036  \\
  $60$  & $0.0568$ & 1.0113  & 1.0070  & 0.0042  \\
  $90$  & $0.0828$ & 1.0164  & 1.0115  & 0.0048  \\
 $120$  & $0.1074$ & 1.0213  & 1.0158  & 0.0054  \\
\end{tabular} 
\end{ruledtabular}
 \end{table}

Second, we show that the relative volume change caused by the Poynting effect (i.e., the normal stress difference $N_{1} \simeq G\gamma^{2}$ under shear deformation) is negligible.
The relative volume change is written as
\begin{equation}
\frac{\Delta V_\mathrm{Poy}}{V}=\frac{N_{1}}{K} \simeq \frac{G\gamma^{2}}{K} < 2 \times 10^{-10},
\end{equation}
where the applied strain is $\gamma =0.01$, and the shear and bulk moduli are $G<40$~kPa and $K>2$~GPa, respectively.
Thus, the volume change due to the Poynting effect is negligible compared to the volume change due to the free thermal expansion ($\Delta V/V\sim 10^{-3}$).
Similarly, other volume changes associated with internal pressure changes during shear deformation are negligible because of a sufficiently large bulk modulus ($K>2$~GPa).\\

In conclusion, the free thermal expansion is dominant, and the relative volume change is negligibly small ($\Delta V/V\sim 10^{-3}$).
Thus, we conclude that the experimental condition can be regarded as a constant-volume condition.
We provide a detailed estimation of the contribution to entropy elasticity by the small volume change in Appendix~\ref{App:Error}.

\section{Contribution to Entropy Elasticity by Small Volume Changes}
\label{App:Error}

In this appendix, we demonstrate that the small volume change due to temperature change (i.e., $\Delta V/V\sim 10^{-3}$ calculated in Appendix~\ref{App:Volume}) has a negligible effect on the analysis of entropy elasticity.
We consider a \textit{compressible} elastic body with an applied shear strain $\gamma$ at temperature $T$ and external pressure $P$.
The derivative of the Helmholtz free energy $F$ is given by \cite{LandauLifshitz}
\begin{equation}
dF = -SdT -PdV + V\sigma d\gamma,
\label{app:eq:dF}
\end{equation}
where $S$, $V$, and $\sigma$ are the entropy, volume, and shear stress, respectively.
We assume that the control parameters are $T$ and $\gamma$ (with constant $P$).
The infinitesimal volume change is
\begin{equation}
dV = V(T,\gamma)\alpha(T,\gamma) dT + \frac{\partial V(T,\gamma)}{\partial \gamma} d \gamma,
\label{app:eq:dV}
\end{equation}
where
\begin{equation}
\alpha (T,\gamma) \equiv \frac{1}{V(T,\gamma)} \frac{\partial V(T,\gamma)}{\partial T}
\end{equation}
is the thermal expansion coefficient \cite{LandauLifshitz}.
Substituting Eq.~(\ref{app:eq:dV}) into Eq.~(\ref{app:eq:dF}), we have
\begin{equation}
dF = -\left(S+PV\alpha \right)dT
+ \left(V \sigma - P\frac{\partial V}{\partial \gamma} \right)
 d\gamma.
\end{equation}
Hereafter, we omit the variables $(T,\gamma)$ because all physical quantities are bivariate functions of $(T,\gamma)$.
By considering $\partial^{2}F/(\partial T\partial \gamma)$, we have the Maxwell relation:
\begin{equation}
 - \frac{\partial}{\partial \gamma} \Bigl(S+ PV\alpha\Bigr)
= \frac{\partial}{\partial T} \left(V \sigma - P\frac{\partial V}{\partial \gamma} \right).
\label{app:eq:Maxwell}
\end{equation}
By using $\partial V/\partial T=V\alpha$ and $\partial P/\partial \gamma=0$ (because $P$ is a constant), we rewrite Eq.~(\ref{app:eq:Maxwell}) as
\begin{equation}
\begin{split}
-\frac{\partial S}{\partial \gamma}
& = \frac{\partial}{\partial T} \left(V \sigma - P\frac{\partial V}{\partial \gamma} \right)
+\frac{\partial}{\partial \gamma} \Bigl( PV\alpha\Bigr)\\
& = V\frac{\partial\sigma}{\partial T} + V\alpha\sigma -P\frac{\partial^{2} V}{\partial T\partial \gamma}
+P\frac{\partial}{\partial \gamma} \Bigl(V\alpha\Bigr)\\
& = V\frac{\partial\sigma}{\partial T} + V\alpha\sigma .
\end{split}
\end{equation}
Therefore, the entropy contribution to the stress is
\begin{equation}
\begin{split}
 \sigma_{S} (T,\gamma)
 \equiv -\frac{T}{V}\frac{\partial S}{\partial \gamma} (T,\gamma)
 = T\frac{\partial\sigma}{\partial T} (T,\gamma)
 + T\alpha\sigma .
\end{split}
\label{app:eq:sigma-vantHoff}
\end{equation}
In Eq.~(\ref{app:eq:sigma-vantHoff}), an additional term $T\alpha\sigma$ is added to Eq.~(\ref{eq:sigma-vantHoff}), as a result of taking into account the volume change.\\

We compare $T\alpha\sigma$ with $T\partial\sigma/\partial T$ in Eq.~(\ref{app:eq:sigma-vantHoff}).
The ratio is
\begin{equation}
\frac{T\alpha\sigma}{T\left(\frac{\partial\sigma}{\partial T}\right)}
\sim
\alpha\Delta T \frac{\sigma}{\Delta \sigma}
\sim 10^{-2},
\label{app:eq:ratio}
\end{equation}
where $\alpha\Delta T \sim \Delta V/V\sim 10^{-3}$ (see Appendix~\ref{App:Volume})
and $\sigma/\Delta \sigma \sim 10$ (see Fig.~\ref{fig:linear}(a)).
Here, $m\sim n$ means that the quantities $m$ and $n$ have the same order of magnitude.
Equation~(\ref{app:eq:ratio}) shows that the additional term $T\alpha\sigma$ caused by the volume change only affects the entropy elasticity insignificantly (by a few percent).\\

We note that the negligible effect of volume change on the temperature dependence of the shear stress (and shear modulus) of the gel is consistent with the absence of thermoelastic inversion, as shown in Fig.~\ref{fig:linear}(a).
Here, the thermoelastic inversion is an inversion of the temperature dependence of stress in the low-strain region caused by thermal expansion, and is observed in various synthetic and natural rubbers \cite{Flory1953,Anthony1942}.

\begin{figure}[t!]
\centering
\includegraphics[width=7.6cm]{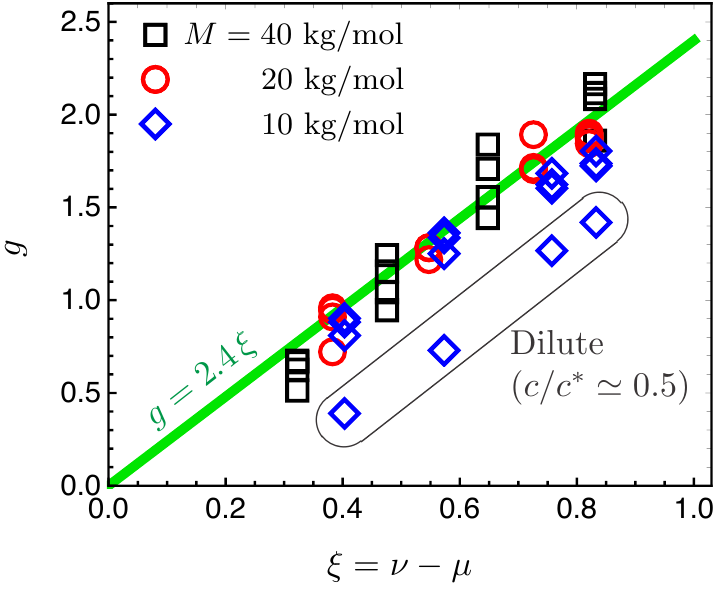}
\caption{
Comparison between the experimentally obtained $g=g(p,c/c^{*})$ [defined in Eq.~(\ref{eq:main-result})] and the dimensionless structure parameter $\xi=\nu-\mu$.
Here, $\nu$ and $\mu$ are the numbers per precursor of elastically effective chains and crosslinks, respectively.
We calculate $\nu$, $\mu$, and $\xi$ using the Bethe approximation with $p$ estimated from Eq.~(\ref{eq:p}).
We determine $g$ from the data in Fig.~\ref{fig:universal}(a) with Eq.~(\ref{eq:main-result}). 
Each symbol represents one sample that is characterized by the molar mass $M$, concentration $c$, and connectivity $p$.
The blue diamonds, red circles, and black squares represent $M=10$, $20$, and $40$ kg/mol, respectively.
The green line represents $g=2.4\,\xi$.
The four blue diamonds enclosed by the gray curve represent the samples with the lowest normalized polymer concentrations ($c/c^* \simeq 0.5$) in this experiment.
They deviate from the green line because they have many elastically ineffective connections such as the intramolecular bonds and loops \cite{Sakai2016, Zhong2016, Yoshikawa2019}, causing the overestimation of $\xi$.
}
\label{fig:xi}
\end{figure}

\section{Bethe approximation to calculate the structural parameters}
\label{App:Bethe-Approximation}

The dimensionless structural parameters $\nu$, $\mu$, and $\xi$ cannot be experimentally observed but can be theoretically estimated as a function of connectivity $p$ by using the Bethe approximation \cite{Macosko1976,Miller1976} (also called the tree approximation \cite{deGennes1979} and the mean-field approximation \cite{Rubinstein2003}).
The number per precursor of the elastically effective cycles $\xi$ is obtained by $\xi=\nu-\mu$, 
where $\nu$ and $\mu$ are the numbers per precursor of elastically effective chains and crosslinks, respectively.
Here, the elastically effective chain is defined as a chain whose ends both connect to crosslinks.\\

In the network structure formed by two kinds of tetra-functional precursors ($\mathrm{A}_{4}$ or $\mathrm{B}_{4}$), the Bethe approximation assumes that the probability that one arm of $\mathrm{A}_{4}$ or $\mathrm{B}_{4}$ does not connect to an infinite-sized network [$P(F^\mathrm{out}_\mathrm{A})$ or $P(F^\mathrm{out}_\mathrm{B})$, respectively] as \cite{Macosko1976,Miller1976,Yoshikawa2019} 
\begin{equation}
\begin{split}
P(F^\mathrm{out}_\mathrm{A})=p_\mathrm{A}P(F^\mathrm{out}_\mathrm{B})^{3} + 1-p_\mathrm{A},\\
P(F^\mathrm{out}_\mathrm{B})=p_\mathrm{B}P(F^\mathrm{out}_\mathrm{A})^{3} + 1-p_\mathrm{B}.
\label{eq:PFout}
\end{split}
\end{equation}
Here, $p_\mathrm{A}$ and $p_\mathrm{B}$ are the fractions of the reacted A and B groups to all functional groups, respectively.
By using $P(F^\mathrm{out}_\mathrm{A})$ and $P(F^\mathrm{out}_\mathrm{B})$, 
we can calculate the probabilities that $\mathrm{A}_{4}$ or $\mathrm{B}_{4}$ will become an $f$-functional crosslink for $f = 3$ or $4$ [$P(X_{\mathrm{A}f})$ or $P(X_{\mathrm{B}f})$, respectively] as
\begin{equation}
\begin{split}
P(X_{\mathrm{A}3})&=4P(F^\mathrm{out}_\mathrm{A})[1-P(F^\mathrm{out}_\mathrm{A})]^{3},\\
P(X_{\mathrm{B}3})&=4P(F^\mathrm{out}_\mathrm{B})[1-P(F^\mathrm{out}_\mathrm{B})]^{3},\\
P(X_{\mathrm{A}4})&=[1-P(F^\mathrm{out}_\mathrm{A})]^{4},\\
P(X_{\mathrm{B}4})&=[1-P(F^\mathrm{out}_\mathrm{B})]^{4}.
\label{eq:PX}
\end{split}
\end{equation}
Note that $\mathrm{A}_{4}$ and $\mathrm{B}_{4}$ cannot be a crosslink for $f = 1~\mathrm{and}~2$.
If the molar ratio of A and B groups is [A] : [B] = $q : 1-q$ ($0 < q < 1$), we have
\begin{equation}
\begin{split}
\nu
&=q \left[\frac{3}{2} P(X_{\mathrm{A}3})+2P(X_{\mathrm{A}4})\right]\\
 & +(1-q)
 \left[\frac{3}{2} P(X_{\mathrm{B}3})+2P(X_{\mathrm{B}4})\right],
\end{split}
\end{equation}
\begin{equation}
\begin{split}
\mu
=q & \left[P(X_{\mathrm{A}3})+P(X_{\mathrm{A}4})\right] \\
& +(1-q)
  \left[ P(X_{\mathrm{B}3})+P(X_{\mathrm{B}4})\right].
\end{split}
\label{eq:numu}
\end{equation}
In this study, we calculate $\nu$ and $\mu$ as a function of connectivity $p = qp_\mathrm{A} + (1-q)p_\mathrm{B}$ \cite{Yoshikawa2019}.\\

We compare $\xi(p)=\nu(p)-\mu(p)$ with $g=g(p,c/c^{*})$ defined in Eq.~(\ref{eq:main-result}).
Figure~\ref{fig:xi} shows $g(p, c/c^{*})\simeq 2.4\,\xi(p)$.
In the semidilute regime ($c/c^{*}>1$), it was reported \cite{Nishi2017, Yoshikawa2019} that the $p$ dependence of $G$ is well reproduced by $G(p) \sim \xi(p)$.
Equation~(\ref{eq:main-result}) with $g(p, c/c^{*})\simeq 2.4\,\xi(p)$ leads to $G(p) \sim \xi(p)$ \cite{Nishi2017,Yoshikawa2019},
which is consistent with the previous experiments \cite{Nishi2017,Yoshikawa2019}.
However, the entropy contribution obtained in this study, $G_{S}=cRg(p)T/M$, is $2.4$ times as large as $G_{S}^{\mathrm{phantom}}=cR\xi(p) T/M$, predicted by the phantom network model \cite{James1953, Rubinstein2003}.

\section{Failure to explain negative energy elasticity of gels by rubber elasticity theory}
\label{App:History-rubber}

We briefly show that the standard rubber elasticity theory of Flory and co-workers \cite{Mark1965, Allen1970, Chen1973, Mark1976, Flory1969} cannot explain the microscopic (molecular) origin of negative energy elasticity in a rubberlike gel. 
This theory considers that the slight energy elasticity in rubber originates from conformational energy change as follows.
We consider the uniaxial tensile deformation of a rubber with an applied stretch ratio $\lambda$ and assume the constant-volume condition.
Following the procedure to derive Eq.~(\ref{eq:sigma-vantHoff}), we have the energy contribution to the tensile stress as
\begin{equation}
\sigma_E(T,\lambda)
=\sigma(T,\lambda)
-T\frac{\partial \sigma}{\partial T}(T,\lambda).
\label{app:eq:SigmaE}
\end{equation}
[Here, we use the symbols $\sigma$ and $\sigma_E$ for tensile deformations, as shown in Fig.~\ref{fig:comparison}(a), while we use the same symbol for shear deformations in the main text.]
On the other hand, the molecular theory of rubber elasticity \cite{Flory1953, James1953} describes the equation of state for rubber elasticity as
\begin{equation}
\sigma(T,\lambda)
=nRT\frac{\langle r^{2}\rangle}{\langle r^{2} \rangle_0}\left(\lambda-\frac{1}{\lambda^{2}}  \right),
\label{app:eq:Gauss}
\end{equation}
where $R$ is the gas constant and $n$ is the molar density of the elastically effective chains (or cycles) for the affine (or phantom) network model.
Here, $\langle r^{2}\rangle \equiv l^2$ and $\langle r^{2} \rangle_0$ are the mean-square linear distance between the crosslinks in the network and the corresponding value for the unperturbed polymer chain, respectively.
If the volume of the rubber is constant, $\langle r^{2}\rangle$ does not depend on temperature $T$.
Substituting Eq. (\ref{app:eq:Gauss}) into Eq. (\ref{app:eq:SigmaE}), we have the formula that describes the energy elasticity of rubbers \cite{Mark1965, Allen1970, Chen1973, Mark1976}:
\begin{equation}
\frac{\sigma_E}{\sigma}
=
\frac{T}{\langle r^2\rangle_0}
\frac{d\langle r^2\rangle_0}{dT}.
\label{app:eq:RIS}
\end{equation}
Equation~(\ref{app:eq:RIS}) is validated in various polymer species including PEG; the value on the left-hand side of Eq.~(\ref{app:eq:RIS}) determined from the thermo-elasticity measurement of unswollen rubbers is consistent with the value on the right-hand side of Eq.~(\ref{app:eq:RIS}) determined from the intrinsic viscosity measurement \cite{Mark1976, Bluestone1974}.
Here, in the intrinsic viscosity measurement, the effect of polymer-solvent interaction is eliminated \cite{Mark1976, Bluestone1974}.
The $T$ dependence of $\langle r^{2} \rangle_0$ in Eq.~(\ref{app:eq:RIS}) is microscopically interpreted to be due to the conformational energy change \cite{Mark1965, Chen1973, Mark1976} based on the rotational isomeric state (RIS) model \cite{Flory1969}.
For example, small and positive energy elasticity $\sigma_{E}/\sigma=0.07\pm0.01$ (green line in Fig.~\ref{fig:energy}) observed in the PEG rubber \cite{Mark1965} is explained by the conformational change around the C--C bond [Fig.~\ref{fig:origin}(a)].\\

On the other hand, in polymer gel elasticity, most previous studies \cite{Mark1977, Patel1992, Akagi2013, Zhong2016} implicitly postulated $\langle r^{2} \rangle_0=\langle r^{2} \rangle$ in Eq.~(\ref{app:eq:RIS}) and neglected energy elasticity.
Even if we do not make such a postulation, the classical rubber elasticity theory of Flory and co-workers \cite{Mark1965, Allen1970, Chen1973, Mark1976,Flory1969} fails to explain the energy elasticity in polymer gels.
The energy elasticity in the PEG rubber ($\sigma_{E}/\sigma=0.07\pm0.01$) is considerably different from that in the PEG gel:
large, negative ($\sigma_{E}/\sigma=G_E/G=-1.81\pm 0.04$ at the maximum),
and strongly concentration-dependent, as shown in Fig.~\ref{fig:energy}.

\section{Mixing and elastic free energies at constant volume}
\label{App:Thermodynamics}

We show that the Helmholtz free energy of mixing ($F_{\mathrm{mix}}$) is irrelevant to the origin of the negative energy elasticity.
To define the energy elasticity ($\sigma_S$ and $G_E$), it is necessary that the volume of the system $V$ does not change with shear deformation.
(This condition is satisfied in this study. See Appendix~\ref{App:Volume}.)
The Helmholtz free energy $F$ of an incompressible polymer gel consists of two separate contributions as \cite{Flory1953}
\begin{equation}
F\left(T,\gamma\right)
=F_{\mathrm{mix}}\left(T\right)
+F_{\mathrm{el}}\left(T,\gamma\right),
\label{app:eq:Ftot}
\end{equation}
where $F_{\mathrm{mix}}$ and $F_{\mathrm{el}}$ are the mixing and elastic free energies, respectively.
Here, $F_{\mathrm{mix}}$ is independent of the applied shear strain $\gamma$ because the volume $V$ does not change with deformation.
In an isothermal process, the shear stress and modulus are related to the free energy as
\begin{equation}
\sigma\left(T,\gamma\right)
\equiv
\frac{1}{V}
\frac{\partial F\left(T,\gamma\right)}{\partial \gamma}
=
\frac{1}{V}
\frac{\partial F_{\mathrm{el}}\left(T,\gamma\right)}{\partial \gamma},
\label{app:eq:Sigma-Fel}
\end{equation}
and
\begin{equation}
G\left(T\right)
\equiv 
\left. \frac{1}{V}\frac{\partial^{2} F}{\partial \gamma^{2}} (T,\gamma) \right|_{\gamma =0}
=
\left. \frac{1}{V}\frac{\partial^{2} F_{\mathrm{el}}}{\partial \gamma^{2}} (T,\gamma) \right|_{\gamma =0}.
\label{app:eq:G-Fel}
\end{equation}
Equations~(\ref{app:eq:Sigma-Fel}) and (\ref{app:eq:G-Fel}) show that $F_{\mathrm{mix}}$ does not contribute to the stress and modulus.
Thus, the models focusing on $F_{\mathrm{mix}}$ of swollen rubbers and gels (such as the Flory-Huggins theory \cite{Flory1953,deGennes1979,Rubinstein2003}) cannot explain the microscopic origin of the negative energy elasticity.

\end{document}